\documentclass[submission, Phys]{SciPost}
\usepackage{amsmath,cases}
\usepackage[utf8]{inputenc}
\usepackage[T1]{fontenc}
\usepackage{lmodern}
\usepackage{graphicx}
\usepackage{amssymb}
\usepackage{gensymb}
\usepackage{bm}
\usepackage{tabularx}
\usepackage{mathtools}
\usepackage{microtype}
\usepackage{siunitx}
\usepackage[version=4]{mhchem}
\usepackage{bbm}
\usepackage{braket}
\usepackage{mathtools}
\usepackage[normalem]{ulem}
\usepackage{layouts}
\usepackage{float}
\usepackage{textgreek}
\usepackage{orcidlink}
\usepackage{todonotes}

\hypersetup{colorlinks}

\newcommand{\comment}[1]{\stepcounter{CommentNumber}\belowpdfbookmark{#1}{\arabic{CommentNumber}}}
\newcounter{CommentNumber}
\begin{document}

\begin{center}{\Large \textbf{
    Probing Cooper pair momentum by quasiparticle steering with planar Josephson junctions
}}\end{center}

\begin{center}
    Isidora Araya Day$^{1,2,3,*}$, Anton R. Akhmerov$^{2,\dagger}$,
    and Antonio R. L. Manesco$^{2,4,\ddagger}$
\end{center}

\begin{center}
    $^{1}$QuTech, Delft University of Technology, Delft 2600 GA, The Netherlands\\
    $^{2}$Kavli Institute of Nanoscience, Delft University of Technology,
    P.O. Box 4056, 2600 GA Delft, The Netherlands\\
    $^{3}$Donostia International Physics Center, P. Manuel de Lardizabal 4,
    20018 Donostia-San Sebastian, Spain\\
    $^{4}$Center for Quantum Devices, Niels Bohr Institute, University of Copenhagen,
    DK-2100 Copenhagen, Denmark\\[4pt]
    $^{*}$\href{mailto:isidora@araya.day}{isidora@araya.day}\\
    $^{\dagger}$\href{mailto:qsteering@antonakhmerov.org}{qsteering@antonakhmerov.org}\\
    $^{\ddagger}$\href{mailto:qsteering@antoniomanesco.org}{qsteering@antoniomanesco.org}
\end{center}

\begin{center}
    July 09, 2026
\end{center}

\section*{Abstract}
The Cooper pair momentum in a superconductor is associated with a phase gradient of the superconducting order parameter.
In general, this momentum is small compared to the Fermi momentum, which makes it challenging to measure.
Josephson junctions, however, enable the creation of large phase gradients and transfer of the Cooper pair momentum to quasiparticles via Andreev reflection.
In this work we demonstrate that Andreev bound states propagating along ballistic planar Josephson junctions eject into an adjacent normal region at a phase-controlled angle that scales as $\Theta \sim \sqrt{\Delta/\mu}$, where $\Delta$ is the superconducting gap and $\mu$ is the chemical potential.
This angle parametrically exceeds the conventional Cooper pair momentum scale $\Delta/\mu$, and thus this phenomenon is sizeable even within the Andreev approximation regime $\Delta/\mu \ll 1$.
Our results establish phase-controlled quasiparticle ejection as a kinematic probe of condensate momentum transfer: unlike existing probes that detect the Doppler energy shift, the signal appears as a momentum-space deflection of emitted quasiparticles.

\section{Introduction}

\comment{Phase gradient in superconductors is the Cooper pair momentum.}
In a moving superconducting condensate every Cooper pair is formed by electrons with wavevectors $\bm{k}$ and $-\bm{k} + \bm{q}$, where $\bm{q}=\nabla \phi$ is the Cooper-pair wavevector and $\hbar\bm{q}$ is the Cooper-pair momentum~\cite{Bardeen57}.
Quasiparticles approaching the superconductor reflect through Andreev processes, which transfer the condensate momentum $\hbar\bm{q}$ to the reflected electrons or holes~\cite{Tkachov2004}.
This momentum mismatch, known as a Doppler shift of the Andreev states, is visible in tunneling spectra~\cite{Fogelstrom97} and nonlocal conductance~\cite{Rohlfing09,Lykkegaard25,Dahl26}.
The accompanying nonreciprocity was recently utilized as one of the mechanisms enabling the Josephson diode effect~\cite{Davydova22,Davydova24}.
These probes detect the condensate momentum indirectly through changes in quasiparticle energies.
Here we instead ask whether the same momentum transfer can be detected kinematically, through the direction of quasiparticle motion after leaving the junction.

\comment{The phase gradient is limited by the inverse coherence length, and therefore the Cooper pair momentum is small in the Andreev approximation regime.}
Because the supercurrent in a superconductor must be lower than its critical current, the Cooper-pair wavevector cannot exceed the inverse coherence length $\xi^{-1} = \Delta / \hbar v_F$, where $\Delta$ is the superconducting gap and $v_F$ is the Fermi velocity.
For the condensate momentum to influence quasiparticle trajectories, the Cooper-pair wavevector must be comparable to the Fermi wavevector $k_F = 2\mu / \hbar v_F$ of a parabolic band, where $\mu$ is the chemical potential.
This ratio $\xi^{-1}/k_F$ between the two is also the small parameter $\Delta/\mu$ that controls the validity of the Andreev approximation.
Therefore, without careful control of the normal state band structure, the Doppler shift of Andreev states is a small effect.

\comment{A ballistic planar Josephson junction hosts an Andreev band, which presents several interesting phenomena}
The phase gradient becomes large at a Josephson junction, where the superconducting order parameter is locally suppressed.
In a clean and sufficiently extended planar junction, the Andreev bound states hybridize along the junction and form dispersive Andreev bands~\cite{Tam2023a}.
Their energies then depend on the momentum parallel to the junction interface, giving the modes a finite velocity along the normal channel.
References~\cite{Tam2023a,Tam2023b} showed that the large momentum separation between counter-propagating Andreev modes strongly suppresses backscattering, leading to robust transport along ballistic planar Josephson junctions.
In earlier work, we extended this phenomenon to multiterminal junctions, where the Fermi-sea topology protects the chiral propagation of Andreev modes between normal terminals~\cite{ArayaDay2025}.
We showed that the momentum structure of the Andreev bands allows to control their propagation and enables large angle scattering between normal terminals despite the absence of normal reflection.
\begin{figure}[!b]
    \centering
    \includegraphics[width=\textwidth]{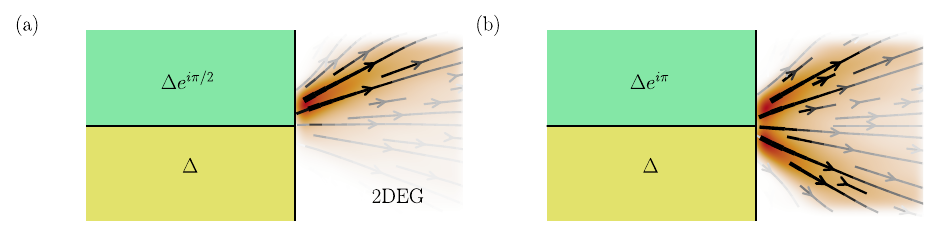}
    \caption{
        Electron current density ejected from a planar Josephson junction into a normal region at subgap energies.
        (a) For a generic phase difference, here $\phi=\pi/2$, quasiparticles are ejected into the normal region at a finite angle.
        (b) At a time-reversal-symmetric phase, here $\phi=\pi$, the ejection is symmetric and the average angle vanishes.
        The arrows show the local electron-current density.
        We use $E = \Delta [1 + \cos(\phi/2)] / 2$, $\Delta = 0.1$, $\mu=0.5$, and $t=1$ for both panels.
    }
    \label{fig:fig1}
\end{figure}

\comment{We show that in two-terminal junctions electrons are ejected with an angle due to the finite momentum of Cooper pairs in the JJ.}
Motivated by the unusual properties of ballistic Josephson junctions, we ask how the superconducting condensate shapes the ejection of quasiparticles out of the junction.
We address this question by studying the scattering of a propagating Andreev bound state at the interface between two leads: a semi-infinite SNS junction with a short distance between the superconductors---a short junction---and a semi-infinite normal region.
Our central result is shown in Fig.~\ref{fig:fig1}: subgap quasiparticles in planar junctions eject at a finite and phase-controlled angle.
In the rest of this work, we demonstrate that this angle scales as $\sqrt{\Delta/\mu}$, which is parametrically larger than the angle limited by the Cooper pair momentum scale, $\sim \Delta/\mu$, common to the currently known manifestations of the Doppler shift.
This larger scale is possible because the ABS forms through repeated Andreev reflections from the phase-biased condensate, rather than through a single reflection event.
We propose as well the detection of this phenomenon using electron optics techniques.

\section{Current ejection from a planar Josephson junction}

\comment{A phase-biased short junction gives Andreev modes a large transverse momentum.}
We consider a planar Josephson junction that is extended along $x$ and has a short separation between the superconductors across $y$:
\begin{equation}
    H_{\textrm{BdG}} = \left(\frac{\hbar^2 k^2}{2m} - \mu\right)\tau_z + \Delta \left[\cos(\phi/2) \tau_x + \sin(\phi/2) \text{sgn}(y) \tau_y\right],
\end{equation}
where $k = \lvert \bm{k} \rvert$ is the magnitude of the wavevector, $\mu$ is the chemical potential of the normal region, $\Delta$ is the proximity-induced superconducting gap, and $\tau_i$ are Pauli matrices acting on the particle-hole degrees of freedom.
The superconducting phase difference $\phi$ is applied across the junction along the $y$ direction.
We choose the phase convention $0\leq\phi\leq\pi$, so that $\sin(\phi/2)$ and $\cos(\phi/2)$ are nonnegative.
For a finite phase difference $\phi$, the junction hosts propagating Andreev modes with energies in the windows $[\Delta \cos(\phi/2), \Delta]$ and $[-\Delta, -\Delta \cos(\phi/2)]$~\cite{Tam2023a}.

\comment{The perturbative ABS momentum sets the ABS-angle scale.}
We estimate the transverse and longitudinal momentum $k_y$ and $k_x$ of the propagating Andreev modes inside the junction using a perturbative expansion beyond the Andreev approximation $\Delta \ll \mu$.
First, we linearize the longitudinal dispersion around the Fermi wavevector $k_F$~\cite{Tam2023a} and we write the effective BdG Hamiltonian as
\begin{equation}
\label{eq:linearized-hamiltonian}
H_{\textrm{BdG}}^{\textrm{eff}} = \left[ak_y^2 + v_x \delta k_x\right]\tau_z + \Delta \cos(\phi/2)\tau_x + \Delta \sin(\phi/2)\operatorname{sgn}(y)\tau_y,
\end{equation}
where $a=\hbar^2/(2m)$ is the kinetic coefficient and $v_x$ is the Fermi velocity along the junction.
We observe that introducing the dimensionless variables
\begin{equation}
\tilde{k}_y=k_y\sqrt{\frac{a}{\Delta}},
\qquad
\tilde{y}=y\sqrt{\frac{\Delta}{a}},
\qquad
\widetilde{\delta k_x}=\frac{v_x\delta k_x}{\Delta},
\qquad
\tilde{E}=\frac{E}{\Delta},
\end{equation}
and dividing Eq.~\eqref{eq:linearized-hamiltonian} by $\Delta$ gives a dimensionless Hamiltonian that depends only on the phase difference $\phi$ and the dimensionless longitudinal momentum $\widetilde{\delta k_x}$:
\begin{equation}
\tilde{H}_{\textrm{BdG}}^{\textrm{eff}} = \left[\tilde{k}_y^2 + \widetilde{\delta k_x}\right]\tau_z + \cos(\phi/2)\tau_x + \sin(\phi/2)\operatorname{sgn}(\tilde{y})\tau_y.
\end{equation}
The rescaled ABS problem therefore has a transverse wavevector of order one in units of $\sqrt{\Delta/a}$.
Comparing this transverse wavevector to the longitudinal Fermi wavevector in the junction, of order $\sqrt{\mu/a}$ due to the parabolic dispersion of the normal state, gives the ABS angle scale $\theta \sim \sqrt{\Delta/\mu}$.
The phase-biased condensate then makes the weights of the opposite transverse components unequal, giving the ABS a nonzero average transverse wavevector on this scale.
This scaling is parametrically larger than the Cooper pair momentum scale $\Delta/\mu$ and sets the momentum angle of the Andreev modes inside the junction.

\comment{The Andreev modes have a transverse momentum that is set by the phase difference and the energy, which can be obtained perturbatively.}
To obtain the energy and phase dependence of the transverse momentum, we find perturbative corrections to the Andreev bound state (ABS) momentum beyond the Andreev approximation---$\Delta \ll \mu$.
Following Ref.~\cite{ArayaDay2025}, we use the two Andreev-approximation basis states in the rescaled variables
\begin{equation}
\braket{\tilde{y}|\Psi_\pm^{(0)}}
=
\sqrt{\frac{\sin(\phi/2)}{2\tilde{k}_0}}
\frac{1}{\sqrt{2}}
\begin{pmatrix}
\pm 1\\
1
\end{pmatrix}
\exp\left[\pm i \tilde{k}_0 \tilde{y} - \frac{\sin(\phi/2)}{2\tilde{k}_0}|\tilde{y}|\right],
\end{equation}
where the Nambu spinor is written in the $\tau_z$ (electron-hole) basis.
The two Andreev-approximation basis states carry opposite dimensionless transverse momenta $\tilde{k}_y = \pm \tilde{k}_0$ and form the perturbative ABS ansatz
\begin{equation}
\ket{\psi_{\mathrm{ABS}}} = A_+\ket{\Psi_+^{(0)}} + A_-\ket{\Psi_-^{(0)}},
\end{equation}
where $A_\pm$ are phase and energy-dependent coefficients.
When the energy of the Andreev modes approaches the gap edge, they become more electron-like or hole-like, and the coefficients $A_\pm$ become imbalanced.
At next-to-leading order in the Andreev approximation, the perturbative Hamiltonian in this basis reads
\begin{equation}
\tilde{H}_{\mathrm{ABS}}^{\mathrm{pert}}
=
\cos(\phi/2)\sigma_z
+ \frac{\sin^2(\phi/2)}{2\tilde{k}_0^2}\sigma_x,
\end{equation}
where $\sigma_i$ act in the $\ket{\Psi_\pm^{(0)}}$ subspace.
The positive eigenvalue satisfies
\begin{equation}
\tilde{E}^2 = \cos^2(\phi/2) + \left(\frac{\sin^2(\phi/2)}{2\tilde{k}_0^2}\right)^2,
\label{eq:perturbative-abs-energy}
\end{equation}
while the corresponding eigenvector has $\langle\sigma_z\rangle = \cos(\phi/2)/\tilde{E}$, and therefore acquires a finite transverse momentum in between $\pm \tilde{k}_0$.
We solve for the dimensionless transverse momentum using Eq.~\eqref{eq:perturbative-abs-energy} and find
\begin{equation}
\tilde{k}_0
=
\frac{\sin(\phi/2)}{\sqrt{2}}\,
\left[\tilde{E}^2-\cos^2(\phi/2)\right]^{-1/4}.
\end{equation}

\begin{figure}[!b]
    \centering
    \includegraphics[width=\textwidth]{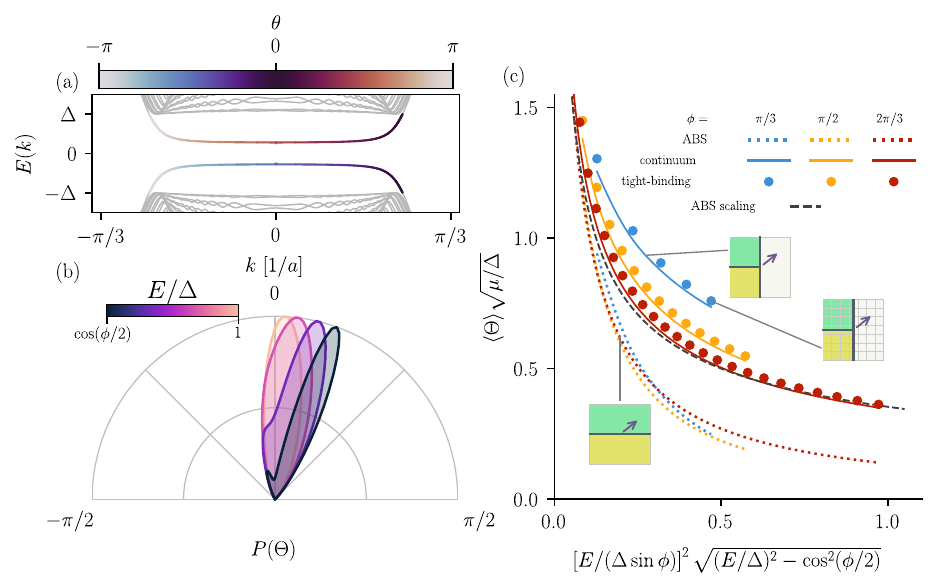}
    \caption{
        Current angle distribution of the ejected electron current density and the Andreev modes inside the junction.
        (a) Dispersion relation of a short planar junction: gray bands provide spectral context, and the colored ABS modes show the angle tilt $\theta$.
        (b) Normalized angular distribution $P(\Theta)$ of the ejected electron current density for a Josephson junction with phase difference $\phi = 2\pi/3$; the colors correspond to different excitation energies $E$.
        (c) Average ejection angle rescaled using the perturbative ABS momentum estimate, plotted as $\langle\Theta\rangle\sqrt{\mu/\Delta}$ against $\left[E/(\Delta\sin\phi)\right]^2\sqrt{(E/\Delta)^2-\cos^2(\phi/2)}$.
        The symbols correspond to tight-binding simulations, the solid lines show the continuum calculation developed in Appendix~\ref{app:analytic-transmitted-electron}, and the dotted and dashed lines show the full and perturbative ABS momentum estimates.
        Different colors show the phase dependence of the ejection angle.
        }
    \label{fig:fig2}
\end{figure}

\comment{We restore the momentum unit and find the ABS transverse momentum.}
Finally, we compute the expectation value of the transverse momentum operator by multiplying the ABS eigenvector polarization $\langle\sigma_z\rangle$ by the transverse momentum $\tilde{k}_0$ and restoring the momentum unit $\sqrt{\Delta/a}$:
\begin{equation}
\label{eq:perturbative-abs-momentum}
\langle k_y\rangle_{\mathrm{ABS}}^{\mathrm{pert}}
=
\sqrt{\frac{\Delta}{a}}\,
\frac{\cos(\phi/2)}{\tilde{E}}\tilde{k}_0
=
\frac{\Delta\sin\phi}{2E}\sqrt{\frac{\Delta}{2a}}\,
\left[E^2/\Delta^2-\cos^2(\phi/2)\right]^{-1/4}.
\end{equation}
This expression has a slow fourth-root divergence in $E - \Delta \cos(\phi/2)$, cut off in a finite-band model once the required transverse kinetic energy becomes comparable to the available normal-state bandwidth.
Away from this cutoff, the transverse ABS momentum is set by $\sqrt{\Delta/a}$ rather than by the much smaller Cooper pair momentum scale.
Figure~\ref{fig:fig2}(a) shows the corresponding tight-binding Andreev bands for $\phi = 5\pi/6$ and $\Delta/\mu=0.2$, with the color scale indicating the angle $\theta=\textrm{arctan}(k_y/k_x)$ of the propagating Andreev modes.
We extract this angle numerically from the longitudinal $J_x^{\textrm{ABS}}$ and transverse $J_y^{\textrm{ABS}}$ electric currents as described in Appendix~\ref{app:tb-angle-extraction}.

\comment{We study how the ABS momentum is transmitted into the normal region.}
At the interface between the junction and a normal region, the Andreev modes are transmitted into pure electron modes~\cite{Tam2023a,ArayaDay2025}.
Additionally, because electrons and holes carry opposite charge and have opposite group velocities with the same momentum, the electric current of an Andreev mode is proportional to the total momentum of its electron and hole components.
Therefore, we study how the phase difference, through the transverse momentum of the Andreev modes, appears in quasiparticle ejection from an SNS-normal junction.
We attach the SNS lead to a semi-infinite normal lead with the same normal Hamiltonian, suppressing ordinary band-mismatch reflection at the interface.
Our goal is to determine whether the ejected quasiparticles inherit the ABS angle scale or whether matching to the normal region substantially changes it.
This question is nontrivial because transparent SNS--normal interfaces transmit Andreev modes into pure electron modes, rather than into electron-hole superpositions~\cite{Tam2023a,ArayaDay2025}, so the wavefunction changes significantly across the interface.
\comment{We solve the continuum scattering problem to obtain the ejection angle.}
We approach this by solving the continuum scattering problem of the Andreev modes at the interface between the SNS junction and the normal region, for which we use the linearized Hamiltonian described by Eq.~\eqref{eq:linearized-hamiltonian}.
We first use the translation invariance along the SNS junction to obtain the full set of modes in the half-plane $x<0$.
It consists of a single propagating Andreev mode and a continuum of evanescent modes that are exponentially decaying in $-x$, and delocalized in $y$.
We then write the transmitted state in the normal region $x>0$ as a superposition of electron modes with different momenta and determine all amplitudes by matching the two sides at the interface.
This construction gives the transmitted electron field and its angular moments for arbitrary phase difference and bias energy; Appendix~\ref{app:analytic-transmitted-electron} presents the full derivation.
We compute the ejection angle from the ratio between the transverse and longitudinal electron currents.
Figure~\ref{fig:fig2}(b) shows the resulting angular distribution, and Fig.~\ref{fig:fig2}(c) shows the corresponding average angle $\langle \Theta \rangle = \arctan\left(j^{e}_y / j^{e}_x\right)$, with $j_y^e/j_x^e=\langle k_y\rangle/k_F$ in the continuum calculation.
The continuum result follows the perturbative ABS momentum estimate rather than the bare Cooper pair momentum scale, confirming the intermediate scaling $\langle \Theta \rangle \sim \sqrt{\Delta/\mu}$.

\comment{We also check the finite-band corrections using tight-binding simulations.}
To test finite-band corrections, we implement the same scattering setup in a tight-binding model using the Kwant package~\cite{Groth_2014}.
We compute the scattering matrix and assign each outgoing electron mode a transmission weight and an emission angle as described in Appendix~\ref{app:tb-angle-extraction}.
In Fig.~\ref{fig:fig2}(c), we compute the tight-binding angle similarly to the continuum angle, as the transmission-weighted average over outgoing modes.
The tight-binding simulations in Fig.~\ref{fig:fig2}(c) provide a finite-band check of the ejection angle and remain in good agreement with the continuum result even at $\Delta/\mu=0.1$.

\section{Experimental detection}

\begin{figure}[h]
    \centering
    \includegraphics[width=0.8\textwidth]{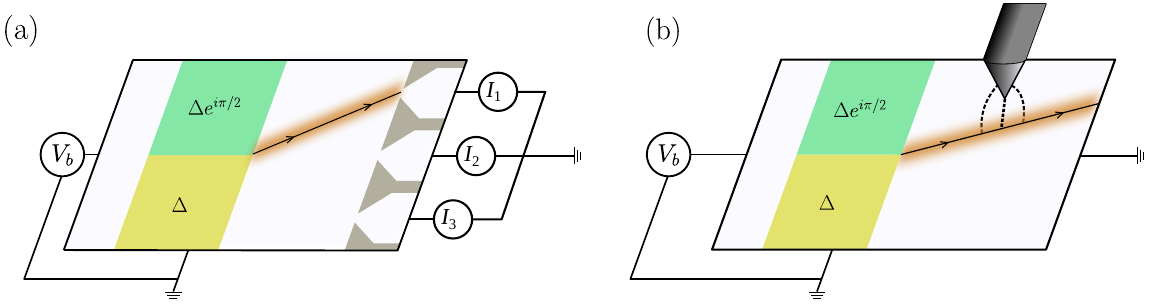}
    \caption{
        Proposed experimental setup to detect the ejection angle of quasiparticles from a planar Josephson junction.
        (a) A short planar Josephson junction is connected to a normal region, where the ejected quasiparticles are detected by a series of quantum point contacts.
        The phase dependence of the nonlocal conductance across each point contact can be used to extract the ejection angle.
        (b) A scanning gate microscopy tip can be used to image the ejected quasiparticle current density.
    }
    \label{fig:fig3}
\end{figure}

\comment{The ejection angle can be measured via electron optics techniques.}
The square-root scaling of the ejection angle with the small parameter $\Delta/\mu$ makes it sizeable even within the Andreev approximation $\Delta/\mu \ll 1$.
Thus, the ejected quasiparticles can be measured with standard electron optics techniques.
For example, the quasiparticle angle distribution in these settings can be detected using quantum point contacts~\cite{Molenkamp90,Shepard92}, with the setup illustrated in Fig.~\ref{fig:fig3}(a).
One can follow the phase dependence of the ejected quasiparticles by measuring the nonlocal conductance $G_i = dI_i/dV_b$ across each point contact $i$ as a function of the phase difference $\phi$ and the bias voltage $V_b$.
A second approach is to directly image the quasiparticle current density via scanning gate microscopy~\cite{Topinka00,Topinka01,Jura09,Bhandari20}, with a setup illustrated in Fig.~\ref{fig:fig3}(b).
Both of these approaches have been used to detect tilted electron jetting due to Fermi surface warping in two-dimensional systems~\cite{Bachmann19,Gold21,InglaAynes23,InglaAynes24}, suggesting that detecting the ejection angle of quasiparticles from a planar Josephson junction is within experimental reach.
To suppress the Hall response in the normal region, it is convenient to bias the Josephson junction with a flux line rather than with an external magnetic field.
It is possible to amplify the ejection angle further via depletion of the normal region in which quasiparticles are ejected, decreasing the Fermi momentum~\cite{Sivan1990,Noguchi1993,Yan2020}.

\comment{Existing proximitized two-dimensional materials have the right gap-to-Fermi-energy scale.}
The material parameter that sets the ejection angle is the gap-to-Fermi-energy ratio $\Delta/\mu$ in the normal region.
For epitaxial InAs--Al devices, reported induced gaps and carrier densities correspond to $\Delta/\mu \sim 10^{-3}$--$10^{-2}$, giving an ejection-angle scale $\sqrt{\Delta/\mu}$ of a few degrees~\cite{Shabani2016,Aghaee2023}.
Graphene devices with lateral superconducting contacts are less suitable because the contacts interrupt the ballistic path and may add substantial scattering.
Top-proximitized graphene devices avoid this constraint and provide a complementary route: Pb or Al islands deposited on graphene induce gaps from a few $10^{-1}\,\mathrm{meV}$ up to nearly $1\,\mathrm{meV}$ over $100\,\mathrm{nm}$--micron length scales~\cite{Natterer16,CortesdelRio24,Trivini25}.
For typical graphene densities $10^{11}$--$10^{12}\,\mathrm{cm}^{-2}$, these values again give $\Delta/\mu$ in the $10^{-3}$--$10^{-2}$ range, and lower-density operation may increase it further.
The recent demonstration of gate-defined Josephson junctions in twisted bilayer graphene~\cite{de_Vries_2021} shows yet another route to realize the proposed setup, since $\Delta \sim 0.1\mathrm{meV}$ and the bandwidth is $\sim 10\mathrm{meV}$, giving $\Delta/\mu \sim 10^{-1}$.
However, in graphene multilayers the Fermi surface is not parabolic, and the ejection angle may be affected by trigonal warping~\cite{Gold21,InglaAynes23,InglaAynes24}.
The magnetic stray field generated by the supercurrent produces a much smaller angular deflection than the ejection effect: estimating one Josephson mode per $k_F^{-1}$ of junction length gives a sheet current $K \sim k_F e\Delta/\hbar$ and a self-field $B \sim \mu_0 K/2$, which bends a quasiparticle trajectory by less than $10^{-4}\,\mathrm{rad}$ over a $1\,\mu\mathrm{m}$ normal region.

\section{Conclusion}

We demonstrated that quasiparticles from a short planar Josephson junction are ejected at a phase and bias-dependent angle.
The ejection angle scales as $\sqrt{\Delta/\mu}$, substantially exceeding the scale set by the Cooper pair momentum $\sim \Delta/\mu$.
The enhancement arises because repeated Andreev reflections from a phase-biased condensate allow the ABS to acquire a momentum larger than that enabled by a single Andreev reflection.
Our analysis relies on ballistic transport of Andreev modes along a planar Josephson junction and on the absence of normal reflection at the junction-normal interface, conditions favored by a short normal region in the junction and a proximity-induced superconducting gap.
Because the effect persists even within the Andreev approximation $\Delta \ll \mu$, experiments using quantum point contacts or scanning gate microscopy are candidates to resolve the phenomenon.

\section*{Acknowledgements}

We acknowledge useful discussions with Josep Ingla-Aynés, Leonid Glazman, and Carlo Beenakker.
I.~A.~D. acknowledges financial support from the Netherlands Organization for Scientific Research (NWO/OCW) as part of the Frontiers of Nanoscience program as well as funding from the European Research Council (ERC) Consolidator grant under grant agreement No. 101042707 (TOPOMORPH).
A.~R.~A. acknowledges funding from an NWO VIDI grant 016.Vidi.189.180.
A.~L.~R.~M. acknowledges the funding from the European Research Council (Grant Agreement No.~856526).

\section*{Author contributions}

A.~R.~L.~M formulated the research idea.
All three authors jointly carried out all other aspects of the work.

\section*{AI disclosure}

The authors used GPT-5.4 and GPT-5.5 to assist with writing and reviewing code and manuscript text.
In particular, the numerical solver for the equations presented in Appendix~\ref{app:analytic-transmitted-electron} was written using GPT-5.5.
All authors reviewed the AI-assisted outputs and take full responsibility for the content of this work.

\section*{Data and code availability}

All code and data used to produce the reported results are available on Zenodo~\cite{ArayaDay_2026}.
The simulations and figures used \texttt{Kwant}~\cite{Groth_2014}, \texttt{Matplotlib}~\cite{Hunter:2007}, \texttt{NumPy}~\cite{2020NumPy-Array}, \texttt{SciPy}~\cite{2020SciPy-NMeth}, and \texttt{SymPy}~\cite{10.7717/peerj-cs.103}.

\bibliographystyle{SciPost_bibstyle}
\bibliography{bibliography}

\appendix
\section{Tight-binding simulation and angle extraction}
\label{app:tb-angle-extraction}

\comment{We introduce the minimal tight-binding BdG model used to generate Fig.~\ref{fig:fig2}.}
We use a square-lattice BdG model whose tight-binding Hamiltonian reads
\begin{equation}
H =
\sum_i c_i^\dagger
\left[
(2t_x+2t_y-\mu)\tau_z
\;+\;
\Delta_i\bigl(\cos\phi_i\,\tau_x+\sin\phi_i\,\tau_y\bigr)
\right] c_i
-\sum_{\langle i,j\rangle} c_i^\dagger t_{ij}\tau_z c_j
\;+\;\mathrm{h.c.},
\end{equation}
where $c_i=(c_{i,e},c_{i,h})^T$ is the Nambu spinor on site $i$ for electrons and holes, and $t_{ij}$ equals $t_x$ or $t_y$ on horizontal or vertical nearest-neighbor bonds.
We set $\Delta_i=0$ in normal regions, and a finite and constant $\Delta_i$ in superconducting ones, with the superconducting phase $\phi_i$ varying across the junction.

\comment{We first use this model to characterize the propagating Andreev modes inside the junction and assign the angle shown in Fig.~\ref{fig:fig2}(a) from their longitudinal and transverse currents.}
To compute the dispersion relation and momentum orientation of the Andreev bands shown in Fig.~\ref{fig:fig2}(a), we solve the short-junction tight-binding half-plane problem directly.
The gray bands in Fig.~\ref{fig:fig2}(a) are obtained from a finite-width Kwant lead only as spectral context; the colored ABS angle is computed from the semi-analytic solution.
For each energy, we find the longitudinal momentum $k_x$ of the propagating ABS without constructing a wide finite system.
At fixed $k_x$, the superconducting half-planes have two decaying roots on each side of the junction.
We form the four-component matching matrix from the tight-binding equations on the two sites adjacent to the junction and find the $k_x$ that minimizes its smallest singular value.
The corresponding right singular vector gives the amplitudes of the four decaying solutions.
We then sum the resulting geometric series analytically to obtain the norm and the transverse current $J_y^{\textrm{ABS}}$, while the longitudinal current is
\begin{equation}
J_x^{\textrm{ABS}}=2t_x\sin k_x\,\langle\psi_{\textrm{ABS}}|\psi_{\textrm{ABS}}\rangle .
\end{equation}
The ABS current angle is
\begin{equation}
\theta(k_x,E)=\operatorname{arctan2}\!\left(J_y^{\textrm{ABS}},J_x^{\textrm{ABS}}\right).
\end{equation}

\comment{We then turn the SNS-normal geometry into a scattering problem and assign each outgoing electron mode a transmission weight and an emission angle.}
For the tight-binding ejection points in Fig.~\ref{fig:fig2}(c), we attach a semi-infinite normal lead to the SNS lead and compute the scattering matrix at fixed energy for an incoming right-propagating Andreev mode.
For each outgoing electron mode $n$ in the normal lead, we use the transmission weight
\begin{equation}
w_n = \sum_m |t_{nm}|^2,
\end{equation}
where $m$ labels incoming Andreev modes.
We evaluate the transverse electron current $J_{y,n}$ carried by the outgoing mode on a vertical bond in the normal lead.
Since the propagating lead modes are normalized to unit longitudinal flux, the corresponding emission angle is
\begin{equation}
\Theta_n=\arctan(J_{y,n}).
\end{equation}

\comment{Finally, we obtain the finite-band ejection angle and the tight-binding ABS curve shown in Fig.~\ref{fig:fig2}(c).}
The tight-binding ejection angle shown by the symbols in Fig.~\ref{fig:fig2}(c) is the corresponding transmission-weighted mean,
\begin{equation}
\langle \Theta \rangle = \frac{\sum_n w_n \Theta_n}{\sum_n w_n}.
\end{equation}
This averaging is equivalent to averaging the transverse and longitudinal currents to leading order in the small ejection angles considered in Fig.~\ref{fig:fig2}(c).
The angular distribution in Fig.~\ref{fig:fig2}(b) is obtained from the continuum transmitted wavefunction in Appendix~\ref{app:analytic-transmitted-electron}, rather than from this finite-width tight-binding calculation.

The tight-binding ABS curve in Fig.~\ref{fig:fig2}(c) uses the same semi-analytic ABS solution as Fig.~\ref{fig:fig2}(a).
There we plot $\operatorname{arctan2}(|J_y^{\textrm{ABS}}|,|J_x^{\textrm{ABS}}|)$, rescaled by the same $\sqrt{\mu/\Delta}$ factor as the continuum curves.

\section{Analytic formulation of the transmitted electron}
\label{app:analytic-transmitted-electron}

This appendix contains an analytic formulation of the electronic wavefunction that ejects from the SNS junction into the normal lead.
Because the wavefunction is the result of a scattering process, it is a superposition of the modes in the normal lead whose amplitudes are determined by the scattering problem at the interface.
Therefore, the goal of this appendix is to find the transmission amplitudes from the SNS junction into the normal lead.
We solve this by first finding all the bound state solutions on the SNS side, and then matching them to the normal-side modes at the interface.

\subsection{Problem Definition}

\subsubsection{Geometry and Hamiltonians}

We consider a geometry with an SNS junction occupying the half-plane $x<0$ and a normal lead in the half-plane $x>0$.
Both sides satisfy the BdG equation
\begin{equation}
H_{\mathrm{BdG}}(x,y)\Psi(x,y)=\epsilon \Psi(x,y),
\qquad
|\epsilon|<1,
\end{equation}
where the normal-side Hamiltonian is
\begin{equation}
H_N(\delta k_x)
=
\left[-\partial_y^2+\delta k_x\right]\tau_z,
\qquad x>0,
\end{equation}
and the SNS-side Hamiltonian is
\begin{equation}
H_{\mathrm{SNS}}(\delta k_x,\phi)
=
\left[-\partial_y^2+\delta k_x\right]\tau_z
+
\cos(\phi/2)\,\tau_x
+
\sin(\phi/2)\,\operatorname{sgn}(y)\,\tau_y,
\qquad x<0.
\end{equation}
The Pauli matrices $\tau_{x,y,z}$ are defined in particle-hole space, $\delta k_x$ is the momentum shift away from the Fermi momentum, and $\phi$ is the superconducting phase difference across the junction.
We use the same phase convention as in the main text, $0\leq\phi\leq\pi$, so that $\sin(\phi/2)$ and $\cos(\phi/2)$ are nonnegative.
For simplicity, we consider a step-function phase profile with a jump of $\phi$ across the $x$-axis, so the SNS Hamiltonian is piecewise constant in $y$, and we work in units where $\Delta=1$ and set $v_x=1$.
The dimensionless energy in this appendix is $\epsilon=E/\Delta$.

The normal lead is gapless.
The longitudinal linearization keeps the outgoing electron-like branch of the normal lead, while the SNS junction has a local minigap of $\cos(\phi/2)$ and supports subgap ABS with energies above the minigap.
Therefore we start by defining the normal-side wavefunction and the matching conditions at the interface $x=0$, to then classify the SNS-side modes and find transmission amplitudes.

\subsubsection{Normal-side wavefunction and matching conditions}

We start by writing the outgoing normal-side wavefunction as a superposition of electron modes without any hole component:
\begin{equation}
\Psi_N(x,y)
=
\begin{pmatrix}
\psi_e(x,y)\\
0
\end{pmatrix},
\qquad
\psi_e(x,y)=\int_{\mathbb{R}} \frac{dq}{2\pi}\, t(q)e^{iqy} e^{i \delta k_x(q) x},
\qquad x>0.
\end{equation}
For the electron component, the normal-side equation gives
\begin{equation}
q^2+ \delta k_x(q)=\epsilon,
\qquad\text{so}\qquad
\delta k_x(q)=\epsilon-q^2.
\end{equation}
Setting $x=0$ gives the boundary value
\begin{equation}
\Psi_N(0,y)
=
\begin{pmatrix}
t(y)\\
0
\end{pmatrix},
\qquad
t(y)=\int_{\mathbb{R}} \frac{dq}{2\pi}\, t(q)e^{iqy}.
\end{equation}

At the interface $x=0$, the wavefunction must be continuous, thus:
\begin{equation}
\psi_e^{\mathrm{SNS}}(0,y)=t(y),
\qquad
\psi_h^{\mathrm{SNS}}(0,y)=0,
\end{equation}
giving two sets of matching conditions.
Additionally, because the SNS-side wavefunctions must remain bounded as $x\to-\infty$, we search for solutions with $\operatorname{Im}\delta k_x\le 0$ and proceed to classify the SNS-side bounded subgap solutions of:
\begin{equation}
H_{\mathrm{SNS}}(\delta k_x,\phi)\Phi(y)=\epsilon \Phi(y),
\end{equation}
where $\Phi(y)$ is the $y$-profile of the SNS-side wavefunction for a translationally invariant SNS junction along the $x$-direction.

\subsection{SNS-side Solutions}

\subsubsection{Mode classification}

For a translationally invariant SNS junction, seek a half-plane mode of the form
\begin{equation}
\Phi(y)=
\begin{pmatrix}
u\\
v
\end{pmatrix}
e^{i k_y y},
\end{equation}
where $u$ and $v$ are the electron and hole spinor components, respectively.
In either half-plane the phase profile is constant, so we define $\sigma=\operatorname{sgn}(y)=\pm 1$ and write the half-plane BdG equation as
\begin{equation}
\begin{pmatrix}
k_y^2+\delta k_x-\epsilon & e^{-i\sigma\phi/2}\\
e^{i\sigma\phi/2} & -k_y^2-\delta k_x-\epsilon
\end{pmatrix}
\begin{pmatrix}
u\\
v
\end{pmatrix}
=0.
\label{eq:half-plane-matrix}
\end{equation}
The determinant condition for a nonzero solution becomes
\begin{equation}
    k_y^2+\delta k_x=\pm i\sqrt{1-\epsilon^2},
\label{eq:half-plane-squared}
\end{equation}
where $1-\epsilon^2 >0$ for subgap states.

We search for solutions with $\operatorname{Im}k_y>0$ for $y > 0$ and $\operatorname{Im}k_y<0$ for $y < 0$ to form interface-localized ABS.

As a result, there are four roots for $k_y$:
\begin{equation}
    k_y = \pm \sqrt{\pm i\sqrt{1-\epsilon^2}-\delta k_x},
\label{eq:half-plane-roots}
\end{equation}
out of which we select those that do not grow as $x\to-\infty$ and do not explode away from the interface $y=0$.
We analyze these conditions in detail below.

\begin{figure}[t]
    \centering
    \includegraphics[width=\textwidth]{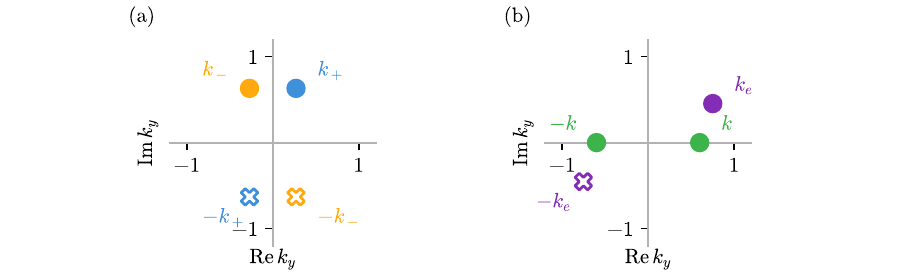}
    \caption{
        Root structure of the bounded subgap solutions of Eq.~\eqref{eq:half-plane-squared} at $\phi = 2\pi/3$ and $\epsilon=0.98$.
        Panel (a) shows the ABS roots in the complex $k_y$ plane.
        Panel (b) shows the mixed family containing a propagating pair $\pm k$ and an evanescent pair $\pm k_e$.
    }
    \label{fig:solution-families-roots}
\end{figure}

\subsubsection{Propagating Andreev bound state}

\paragraph{Decaying roots}

We start by analyzing the type of solutions that decay away from the interface $y=0$ in each half-plane and thus form Andreev bound states.
These are the solutions for which $\operatorname{Im}k_y<0$ for $y<0$ and $\operatorname{Im}k_y>0$ for $y>0$.
Equation \eqref{eq:half-plane-squared} defines two branches of solutions for $k_y^2$ in each half-plane, only one of which can produce a decaying root away from the interface $y=0$.
This is because $e^{ik_y y}\leftrightarrow e^{-ik_y y}$, so the two signs correspond to one decaying and one growing solution in a given half-plane.

Let $k_\pm$ be the decaying roots in the upper half-plane, where the $\pm$ label indicates the branch of the squared equation to which they belong:
\begin{equation}
k_+^2+\delta k_x=i\sqrt{1-\epsilon^2},
\qquad
k_-^2+\delta k_x=-i\sqrt{1-\epsilon^2}.
\end{equation}
Substituting these relations into Eq. \eqref{eq:half-plane-matrix} gives:
\begin{equation}
\begin{pmatrix}
\pm i\sqrt{1-\epsilon^2}-\epsilon & e^{-i\sigma\phi/2}\\
e^{i\sigma\phi/2} & \mp i\sqrt{1-\epsilon^2}-\epsilon
\end{pmatrix}
\begin{pmatrix}
u_\pm\\
v_\pm
\end{pmatrix}
=0,
\end{equation}
whose solutions are the spinors:
\begin{equation}
\chi_\pm^{(\sigma)}
=
\begin{pmatrix}
1\\
e^{i\eta_\pm} e^{i\sigma\phi/2}
\end{pmatrix}.
\label{eq:half-plane-spinors}
\end{equation}
where
\begin{equation}
e^{i\eta_\pm}=\epsilon \mp i\sqrt{1-\epsilon^2}.
\end{equation}
Therefore, the half-plane eigenstates that decay away from the $y=0$ interface are the superpositions
\begin{align}
\Phi^{\mathrm{ABS}}_{>}(y)&=C_+\,\chi_+^{(+)} e^{i k_+ y}+C_-\,\chi_-^{(+)} e^{i k_- y},
\qquad y>0,
\\
\Phi^{\mathrm{ABS}}_{<}(y)&=D_+\,\chi_+^{(-)} e^{-i k_+ y}+D_-\,\chi_-^{(-)} e^{-i k_- y},
\qquad y<0,
\end{align}
where $C_\pm$ and $D_\pm$ are the respective amplitudes.

\paragraph{Matching at $y=0$.}

To find the Andreev bound states, we search for solutions that satisfy the boundary conditions at $y=0$.
\begin{equation}
\Phi^{\mathrm{ABS}}_{>}(0)=\Phi^{\mathrm{ABS}}_{<}(0),
\qquad
\partial_y\Phi^{\mathrm{ABS}}_{>}(0)=\partial_y\Phi^{\mathrm{ABS}}_{<}(0).
\end{equation}
Introducing the convenient parameters
\begin{equation}
\gamma=\frac{k_+}{k_-},
\qquad
z=\frac{e^{i\eta_+}}{e^{i\eta_-}} = \frac{\epsilon - i\sqrt{1-\epsilon^2}}{\epsilon + i\sqrt{1-\epsilon^2}},
\label{eq:gamma-z-def}
\end{equation}
this gives four equations for the four unknown amplitudes $C_\pm$ and $D_\pm$:
\begin{align}
C_+ + C_- - D_+ - D_- &= 0,
\\
z e^{i\phi/2} C_+ + e^{i\phi/2} C_-
- z e^{-i\phi/2} D_+ - e^{-i\phi/2} D_- &= 0,
\\
\gamma C_+ + C_- + \gamma D_+ + D_- &= 0,
\\
z\gamma e^{i\phi/2} C_+ + e^{i\phi/2} C_-
+ z\gamma e^{-i\phi/2} D_+ + e^{-i\phi/2} D_- &= 0,
\end{align}
These four equations form the homogeneous $4\times4$ system
\begin{equation}
M(\gamma;z,\phi)
\begin{pmatrix}
C_+\\
C_-\\
D_+\\
D_-
\end{pmatrix}
=0,
\end{equation}
with
\begin{equation}
M(\gamma;z,\phi)
=
\begin{pmatrix}
1 & 1 & -1 & -1\\
z e^{i\phi/2} & e^{i\phi/2} & -z e^{-i\phi/2} & -e^{-i\phi/2}\\
\gamma & 1 & \gamma & 1\\
z\gamma e^{i\phi/2} & e^{i\phi/2} & z\gamma e^{-i\phi/2} & e^{-i\phi/2}
\end{pmatrix}.
\end{equation}

We find the nontrivial solutions of this system by finding the values of $\gamma$ for which $\det M(\gamma;z,\phi)=0$ and then find the conditions on $z$ and $\phi$ for which these roots yield physical ABS modes:
\begin{equation}
z \sin^2(\phi/2)\,(\gamma-1)^2=(z-1)^2\gamma.
\end{equation}
Replacing $z$ by its definition in Eq. \eqref{eq:gamma-z-def} gives two roots for $\gamma = k_+/k_-$:
\begin{equation}
\gamma_\pm=
e^{\pm 2i\beta},
\qquad
\beta=\arcsin\left(\frac{\sqrt{1-\epsilon^2}}{\sin(\phi/2)}\right).
\label{eq:gamma-roots}
\end{equation}
Combining Eq. \eqref{eq:gamma-roots} with Eq. \eqref{eq:gamma-z-def} and Eq. \eqref{eq:half-plane-squared} gives the value of $k_-^2$ for each root:
\begin{equation}
k_-^2=\frac{2i\sqrt{1-\epsilon^2}}{\gamma^2-1}.
\end{equation}
This fixes $k_-^2$ but not the sign of $k_-$.
We choose the sign so that $\operatorname{Im}k_->0$, and then set $k_+=\gamma k_-$.
With the branch conventions above, only
\begin{equation}
\gamma^{\mathrm{ABS}}=\gamma_-=e^{-2i\beta}
\end{equation}
gives $\operatorname{Im}k_+>0$.
The other algebraic root
$\gamma_+=e^{2i\beta}$ gives $\operatorname{Im}k_+<0$ and therefore grows in one half-plane.
The physical ABS therefore comes from the single root $\gamma^{\mathrm{ABS}}$.
Its longitudinal shift is
\begin{equation}
\delta k_x
=
i\sqrt{1-\epsilon^2}-k_+^2
=
-i\sqrt{1-\epsilon^2}-k_-^2.
\end{equation}

With these branch conventions, the matching problem yields a single physical interface-localized mode together with a corresponding right null vector
\begin{equation}
\begin{pmatrix}
C_+\\
C_-\\
D_+\\
D_-
\end{pmatrix}
\end{equation}
that determines the relative amplitudes of the decaying roots in each half-plane only as a function of $\phi$ and $\epsilon$.
We extract the null vector from a singular-value decomposition of $M$: the right singular vector associated with the smallest singular value is the null-space direction, up to an overall normalization and phase.
Figure~\ref{fig:solution-families-roots}(a) shows the ABS roots for a representative value of $\phi$ and $\epsilon$, while Fig.~\ref{fig:solution-families-profiles}(a) and Fig.~\ref{fig:solution-families-profiles}(b) show the corresponding two-dimensional and transverse profiles.

\subsubsection{Mixed evanescent-propagating modes}

\paragraph{Propagating roots}
A second family of solutions follows from asking for one branch of Eq. \eqref{eq:half-plane-squared} to have real transverse momentum:
\begin{equation}
k_y=\pm k \in\mathbb{R},
\qquad
k>0.
\end{equation}
Then the squared dispersion gives
\begin{equation}
\delta k_x(k)=\pm i\sqrt{1-\epsilon^2}-k^2.
\end{equation}
Since every SNS-side mode carries a factor $e^{i\delta k_x x}$ with $x<0$, boundedness as $x\to-\infty$ requires $\operatorname{Im}\delta k_x\le 0$.
Therefore the physical branch is
\begin{equation}
\delta k_x(k)=-i\sqrt{1-\epsilon^2}-k^2.
\end{equation}
This mode is oscillatory in $y$ and evanescent in $x$, so it cannot serve as the incoming propagating source.
Its role is instead to provide the continuum correction needed to enforce $\psi_h(0,y)=0$ pointwise.

\paragraph{Evanescent roots}
Once $\delta k_x(k)$ is fixed in this way, the other branch of Eq. \eqref{eq:half-plane-squared} gives
\begin{equation}
k_e^2(k)=k^2+2i\sqrt{1-\epsilon^2}.
\label{eq:ke-def}
\end{equation}
We therefore define
\begin{equation}
k_e(k)=\sqrt{k^2+2i\sqrt{1-\epsilon^2}},
\qquad
\gamma(k)=\frac{k_e(k)}{k}.
\end{equation}
Choosing the square root of Eq. \eqref{eq:ke-def} with $\operatorname{Im}k_e(k)>0$ gives the $y$-evanescent companion to the propagating pair $\pm k$.
Thus, for each $k>0$, the local solution space contains one evanescent piece and two propagating pieces in each half-plane: this is the mixed $k/k_e$ family.

Using the spinors defined in Eq. \eqref{eq:half-plane-spinors}, the corresponding half-plane wavefunctions are
\begin{align}
\Phi^{k}_{>}(y)
=&\;
A_e\,\chi_+^{(+)} e^{i k_e y}
+
A_+\,\chi_-^{(+)} e^{i k y}
+
A_-\,\chi_-^{(+)} e^{-i k y},
\qquad y>0,
\\
\Phi^{k}_{<}(y)
=&\;
B_e\,\chi_+^{(-)} e^{-i k_e y}
+
B_+\,\chi_-^{(-)} e^{-i k y}
+
B_-\,\chi_-^{(-)} e^{i k y},
\qquad y<0.
\end{align}
Here the factors $e^{\pm i k y}$ are the propagating pieces, while $e^{i k_e y}$ for $y>0$ and $e^{-i k_e y}$ for $y<0$ are the decaying $y$-evanescent pieces selected by $\operatorname{Im}k_e>0$.
The six amplitudes $A_e,A_\pm,B_e,B_\pm$ are the unknown coefficients of the local solution space for each $k>0$.

\paragraph{Matching at $y=0$.}
The six amplitudes $A_e,A_\pm,B_e,B_\pm$ are unknown coefficients that parameterize the local solution space for each $k>0$, and we find them by imposing the matching conditions at $y=0$.
The continuity of the wavefunction and its derivative at $y=0$ gives the system
\begin{align}
A_e-B_e+A_++A_- - B_+ - B_- &=0,
\\
z\left(A_e e^{i\phi/2}-B_e e^{-i\phi/2}\right)
+
\left(A_++A_-\right)e^{i\phi/2}
-
\left(B_++B_-\right)e^{-i\phi/2}
&=0,
\\
\gamma\left(A_e+B_e\right)+\left(A_+-A_-\right)+\left(B_+-B_-\right)&=0,
\\
z\gamma\left(A_e e^{i\phi/2}+B_e e^{-i\phi/2}\right)
+
\left(A_+-A_-\right)e^{i\phi/2}
+
\left(B_+-B_-\right)e^{-i\phi/2}
&=0.
\end{align}
This $4\times 6$ local system has rank $4$ and nullity $2$, so the solution space is two-dimensional.

\begin{figure}[t]
    \centering
    \includegraphics[width=\textwidth]{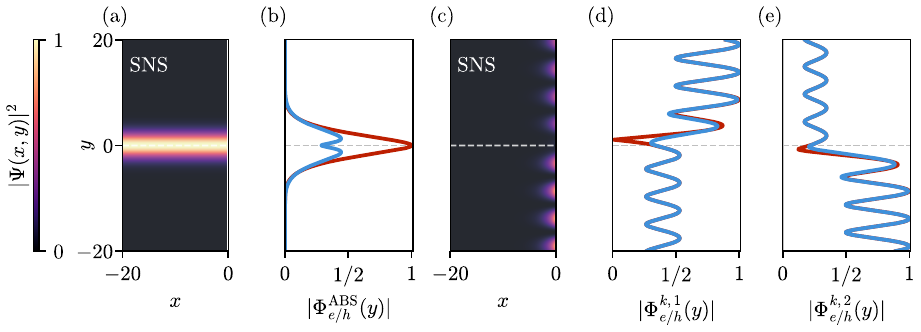}
    \caption{
        Real-space profiles of the bounded subgap solution families shown in Fig.~\ref{fig:solution-families-roots}, for the same parameters $\phi = 2\pi/3$ and $\epsilon=0.98$.
        Panels (a) and (c) show normalized two-dimensional profiles on the SNS side, with the ABS and mixed contributions shown separately.
        Panel (b) shows the corresponding transverse electron and hole profile, arranged as a vertical cut of panel (a).
        Panels (d) and (e) show the transverse profiles of the two independent mixed basis solutions, arranged as vertical cuts associated with panel (c).
    }
    \label{fig:solution-families-profiles}
\end{figure}

\paragraph{Basis Solutions}
To solve for the coefficients we choose a convenient basis of solutions by fixing the evanescent amplitudes $(A_e,B_e)$ in turn.
We choose $(A_e,B_e)=(1,0)$ to get the first basis solution, which gives
\begin{equation}
\begin{aligned}
A_+^1(k)&=-\frac{\gamma(k)z e^{i\phi}-\gamma(k)+z e^{i\phi}-1}{2\bigl(e^{i\phi}-1\bigr)},
\\
A_-^1(k)&=\frac{\gamma(k)z e^{i\phi}-\gamma(k)-z e^{i\phi}+1}{2\bigl(e^{i\phi}-1\bigr)},
\\
B_+^1(k)&=\frac{e^{i\phi}\bigl(\gamma(k)-1\bigr)(z-1)}{2\bigl(e^{i\phi}-1\bigr)},
\\
B_-^1(k)&=-\frac{e^{i\phi}\bigl(\gamma(k)+1\bigr)(z-1)}{2\bigl(e^{i\phi}-1\bigr)}.
\end{aligned}
\label{eq:basis-solution1}
\end{equation}
As a second basis solution, we choose $(A_e,B_e)=(0,1)$, which gives
\begin{equation}
\begin{aligned}
A_+^2(k)&=-\frac{\bigl(\gamma(k)-1\bigr)(z-1)}{2\bigl(e^{i\phi}-1\bigr)},
\\
A_-^2(k)&=\frac{\bigl(\gamma(k)+1\bigr)(z-1)}{2\bigl(e^{i\phi}-1\bigr)},
\\
B_+^2(k)&=\frac{\gamma(k)z-\gamma(k)e^{i\phi}+z-e^{i\phi}}{2\bigl(e^{i\phi}-1\bigr)},
\\
B_-^2(k)&=-\frac{\gamma(k)z-\gamma(k)e^{i\phi}-z+e^{i\phi}}{2\bigl(e^{i\phi}-1\bigr)}.
\end{aligned}
\label{eq:basis-solution2}
\end{equation}
These expressions assume $e^{i\phi}\neq 1$, that is, $\phi\not\equiv 0\;(\mathrm{mod}\;2\pi)$.
The two basis solutions in Eqs.~\eqref{eq:basis-solution1} and \eqref{eq:basis-solution2} determine the coefficient functions entering the continuum kernels, which are fully determined by $\phi$, $\epsilon$, and $k$.
Figure~\ref{fig:solution-families-roots}(b) shows the $k/k_e$ family of roots for a representative value of $\phi$ and $\epsilon$.
The two-dimensional mixed profile is shown in Fig.~\ref{fig:solution-families-profiles}(c), and the real-space profiles of the electron and hole components of the two basis solutions are shown in Figs.~\ref{fig:solution-families-profiles}(d) and \ref{fig:solution-families-profiles}(e).
The real-space asymmetry of the two basis solutions arises for phase differences $\phi\neq 0,\pi$.

The final wavefunction for the mixed family is a superposition of the two basis solutions, which we write as
\begin{equation}
\Phi^{k}(y)
=A_e(k)\,\Phi^{k,1}(y)+B_e(k)\,\Phi^{k,2}(y),
\end{equation}
where $A_e(k)$ and $B_e(k)$ are free coefficients, and $\Phi^{k,1}$ and $\Phi^{k,2}$ are the two basis solutions defined by Eqs.~\eqref{eq:basis-solution1} and \eqref{eq:basis-solution2}:
\begin{align}
\Phi^{k,1}_>(y)
&=
\chi_+^{(+)} e^{i k_e y}
+
A_+^1(k)\,\chi_-^{(+)} e^{i k y}
+
A_-^1(k)\,\chi_-^{(+)} e^{-i k y},
\qquad y>0,
\\
\Phi^{k,1}_<(y)
&=
B_+^1(k)\,\chi_-^{(-)} e^{-i k y}
+
B_-^1(k)\,\chi_-^{(-)} e^{i k y},
\qquad y<0,
\\
\Phi^{k,2}_>(y)
&=
A_+^2(k)\,\chi_-^{(+)} e^{i k y}
+
A_-^2(k)\,\chi_-^{(+)} e^{-i k y},
\qquad y>0,
\\
\Phi^{k,2}_<(y)
&=
\chi_+^{(-)} e^{-i k_e y}
+
B_+^2(k)\,\chi_-^{(-)} e^{-i k y}
+
B_-^2(k)\,\chi_-^{(-)} e^{i k y},
\qquad y<0.
\end{align}

\subsection{Full Half-Line Integral Equations}

In order to find the transmitted electron field in the normal lead, we need to solve the scattering problem at the interface $x=0$ between the SNS junction and the normal lead.
Let $\delta k_x^{\mathrm{ABS}}$ denote the real longitudinal shift of the propagating ABS source mode when that mode exists.
We then assemble the full SNS-side wavefunction as a superposition of the ABS mode $\Phi^{\mathrm{ABS}}$ and the mixed modes $\Phi^{k,1}$ and $\Phi^{k,2}$:
\begin{equation}
\Psi_{\mathrm{SNS}}(x,y)
=
e^{i\delta k_x^{\mathrm{ABS}} x}\,\Phi^{\mathrm{ABS}}(y)
+
\int_0^\infty dk\,
\Big[
A_e(k)\,e^{i\delta k_x(k)x}\Phi^{k,1}(y)
+
B_e(k)\,e^{i\delta k_x(k)x}\Phi^{k,2}(y)
\Big].
\end{equation}
Here $\operatorname{Im}\delta k_x(k)<0$ for the mixed correction modes.

\paragraph{Matching at $x=0$}
Substituting this decomposition into $\psi_h^{\mathrm{SNS}}(0,y)=0$ gives the equations that determine the unknown coefficient functions $A_e(k)$ and $B_e(k)$.
Because the upper and lower half-planes are parameterized differently at $y=0$, we introduce a positive coordinate $s>0$ and write separate equations for $y=s$ and $y=-s$:
\begin{align}
\int_0^\infty dk\,
\Big[
K_{ua}(s,k)\,A_e(k)+K_{ub}(s,k)\,B_e(k)
\Big]
&=
R_u(s),
\\
\int_0^\infty dk\,
\Big[
K_{la}(s,k)\,A_e(k)+K_{lb}(s,k)\,B_e(k)
\Big]
&=
R_l(s),
\end{align}
where the kernels are defined by the hole component of the mixed modes at $x=0$:
\begin{align}
K_{ua}(s,k)
&=
z e^{i k_e(k) s}
+
A_+^1(k) e^{i k s}
+
A_-^1(k) e^{-i k s},
\\
K_{ub}(s,k)
&=
A_+^2(k) e^{i k s}
+
A_-^2(k) e^{-i k s},
\\
K_{la}(s,k)
&=
B_+^1(k) e^{i k s}
+
B_-^1(k) e^{-i k s},
\\
K_{lb}(s,k)
&=
z e^{i k_e(k) s}
+
B_+^2(k) e^{i k s}
+
B_-^2(k) e^{-i k s},
\end{align}
and the right-hand sides are defined by the hole component of the ABS mode at $x=0$:
\begin{equation}
R_u(s)=- z C_+ e^{i k_+ s} - C_- e^{i k_- s},
\qquad
R_l(s)=- z D_+ e^{i k_+ s} - D_- e^{i k_- s}.
\end{equation}

We solve for $A_e(k)$ and $B_e(k)$ by representing the coefficient functions on a finite linear $k$ grid, discretizing the integral equations on a grid of $s$ values, and performing a regularized least-squares fit.
The calculation uses a compact smooth basis for the coefficient functions and evaluates the residual on a denser $s$ grid than the one used for fitting; the published code specifies the grid sizes, cutoffs, basis size, regularization, and residual checks used for each data point.
Thus we define the residual of a trial solution $(A_e(k),B_e(k))$:
\begin{equation}
\mathcal{R}(s)
=
\begin{pmatrix}
\int_0^\infty dk\,[K_{ua} A_e + K_{ub} B_e] - R_u \\
\int_0^\infty dk\,[K_{la} A_e + K_{lb} B_e] - R_l
\end{pmatrix}.
\end{equation}
We find a solution for $(A_e(k),B_e(k))$ by minimizing the relative residual:
\begin{equation}
\frac{\|\mathcal{R}\|_2}{\|(R_u,R_l)\|_2}.
\end{equation}

\subsection{Reconstruction of the Transmitted Electron Field}

\begin{figure}[t]
    \centering
    \includegraphics[width=\textwidth]{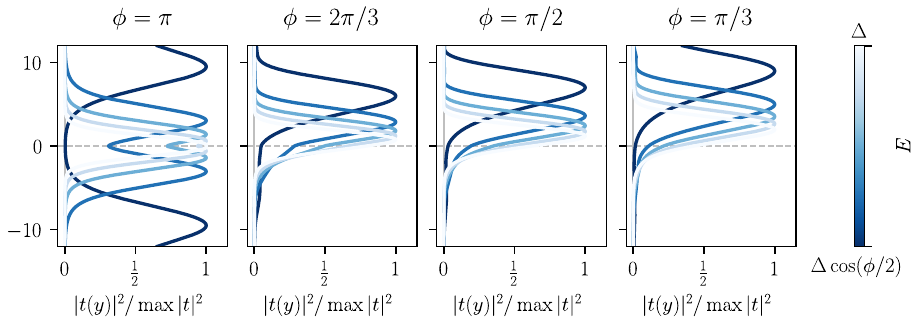}
    \caption{
        Transmitted intensity profiles $|t(y)|^2$ at the SNS--normal interface, obtained from the dense linear-$k$ solve of the half-line integral equations.
        Each panel corresponds to a different phase difference $\phi$, and each curve corresponds to one of five evenly spaced energies between the local minigap $\Delta\cos(\phi/2)$ and the gap edge $\Delta$.
        In every panel, the plotted intensity is normalized by $\max_y |t(y)|^2$.
    }
    \label{fig:transmitted-profile-example}
\end{figure}

Once $A_e(k)$ and $B_e(k)$ are known, the electronic wavefunction in the normal lead follows directly.
Similarly, we define:
\begin{equation}
t_u(s)=t(+s),
\qquad
t_l(s)=t(-s),
\qquad
s>0.
\end{equation}
For $y=+s$,
\begin{multline}
t_u(s)
=
\int_0^\infty dk\,
\Big[
e^{i k_e(k) s} A_e(k)
+
e^{i k s}\big(A_+^1(k) A_e(k) + A_+^2(k) B_e(k)\big)
\\
+
e^{-i k s}\big(A_-^1(k) A_e(k) + A_-^2(k) B_e(k)\big)
\Big]
+
C_+ e^{i k_+ s}
+
C_- e^{i k_- s},
\end{multline}
and for $y=-s$,
\begin{multline}
t_l(s)
=
\int_0^\infty dk\,
\Big[
e^{i k_e(k) s} B_e(k)
+
e^{i k s}\big(B_+^1(k) A_e(k) + B_+^2(k) B_e(k)\big)
\\
+
e^{-i k s}\big(B_-^1(k) A_e(k) + B_-^2(k) B_e(k)\big)
\Big]
+
D_+ e^{i k_+ s}
+
D_- e^{i k_- s}.
\end{multline}
These expressions fully determine the transmitted electronic wavefunction.
Figure~\ref{fig:transmitted-profile-example} shows the transmitted intensity profile $|t(y)|^2$ for several phase differences and several energies within the corresponding ABS window.
In each panel, the intensity is normalized by its maximum over $y$, so the figure should be read as a comparison of profile shapes across phase and energy rather than as an absolute transmission probability.

\subsection{Energy Dependence of the Ejection Direction}

Finally, we obtain the full energy and phase dependence of the transmitted electron field by repeating the above procedure for different values of $\phi$ and $\epsilon$.
The transverse momentum of the transmitted electron field controls its ejection direction.
In the dimensionless units of this appendix, we compute its mean and width as
\begin{equation}
\langle k_y\rangle
=
\frac{\int_{-\infty}^\infty dy\, t^*(y)\left(-i\frac{d}{dy}\right) t(y)}{\int_{-\infty}^\infty dy\, |t(y)|^2},
\qquad
\sigma_{k_y}^2
=
\frac{\int_{-\infty}^\infty dy\, t^*(y)\left(-\frac{d^2}{dy^2}\right) t(y)}{\int_{-\infty}^\infty dy\, |t(y)|^2} - \langle k_y\rangle^2.
\end{equation}
Restoring physical units gives $\Theta=\arctan\left(\sqrt{\Delta/a}\,\langle k_y\rangle/k_F\right)$.

\end{document}